\documentclass[apj]{emulateapj}
\usepackage{amsmath,graphicx,times,bm,url,xspace,color}

\defcitealias{2010MNRAS.405...41G}{G10}
\defcitealias{2012ApJ...761..116B}{BCMR}

\newcommand{\Gre}{\citetalias{2010MNRAS.405...41G}\xspace}
\newcommand{\Bodo}{\citetalias{2012ApJ...761..116B}\xspace}

\newcommand{\Eq}[1]{Equation~(\ref{#1})}

\newcommand{\Fig}[1]{Figure~\ref{#1}}
\newcommand{\Figs}[2]{Figures~\ref{#1} and \ref{#2}}

\newcommand{\Tab}[1]{Table~\ref{#1}}

\usepackage{ulem}

\renewcommand\emph[1]{\textit{#1}}

\newcommand\m{\,\rm m}

\newcommand\tms{\!\times\!}
\newcommand\cdt{\!\cdot\!}

\newcommand\xx{\hat{{\mathbf x}}}
\newcommand\yy{\hat{{\mathbf y}}}
\newcommand\zz{\hat{{\mathbf z}}}

\newcommand\V{\mathbf v}
\newcommand\B{\mathbf B}

\newcommand\U{\mathbf{u}}

\newcommand\mU{\mn{\mathbf{u}}}
\newcommand\mB{\mn{\mathbf{B}}}
\newcommand\EMF{\mbox{\boldmath{${\cal E}$}}}
\newcommand{\OO}{\bm{\Omega}}

\newcommand{\mn}[1]{\overline{#1}}
\newcommand{\rms}[1]{\left<\right.\!#1\!\left.\right>}

\newcommand{\simgt}%
           {\,\hbox{\lower0.35ex\hbox{$\sim$}\llap{\raise0.35ex\hbox{$>$}}}\,}
\newcommand{\simlt}%
           {\,\hbox{\lower0.35ex\hbox{$\sim$}\llap{\raise0.35ex\hbox{$<$}}}\,}

\newcommand\NIII{\textsc{Nirvana-iii}\xspace}

\begin{document}

\title{Dynamo effects in magnetorotational turbulence with finite
  thermal diffusivity}

\author{Oliver Gressel}

\affil{
  NORDITA, KTH Royal Institute of Technology and Stockholm University, 
  Roslagstullsbacken 23, 106 91 Stockholm,   Sweden\\
}

\email{oliver.gressel@nordita.org}



\begin{abstract}
We investigate the saturation level of hydromagnetic turbulence driven
by the magnetorotational instability in the case of vanishing net
flux. Motivated by a recent paper of \citeauthor*{2012ApJ...761..116B},
we here focus on the case of a non-isothermal equation of state with
constant thermal diffusivity.
The central aim of the paper is to complement the previous result with
closure parameters for mean-field dynamo models, and to test the
hypothesis that the dynamo is affected by the mode of heat transport.
We perform computer simulations of local shearing-box models of
stratified accretion disks with approximate treatment of radiative
heat transport, which is modeled via thermal conduction. We study the
effect of varying the (constant) thermal diffusivity, and apply
different vertical boundary conditions.
In the case of impenetrable vertical boundaries, we confirm the
transition from mainly conductive to mainly convective vertical heat
transport below a critical thermal diffusivity. This transition is
however much less dramatic when more natural outflow boundary
conditions are applied. Similarly, the enhancement of magnetic
activity in this case is less pronounced.
Nevertheless, heating via turbulent dissipation determines the
thermodynamic structure of accretion disks, and clearly affects the
properties of the related dynamo. This effect may however have been
overestimated in previous work, and a careful study of the role played
by boundaries will be required.
\end{abstract}



\section{Introduction}
\label{sec:intro}

There are few concepts in classical physics that are equally
fundamental as the conservation of angular momentum. One formidable
consequence of this law is the formation of gaseous accretion disks
around a wide range of astrophysical objects. Yet to explain observed
luminosities based on the release of gravitational binding energy
\citep{2007MNRAS.376.1740K}, a robust mechanism is required to
circumvent the consequences of angular momentum conservation. Being
sufficiently ionised, these discs harbor dynamically important
magnetic fields, which render the disk unstable to a mechanism called
magnetorotational instability \citep[MRI,][]{1998RvMP...70....1B},
releasing energy from the differential rotation and converting it into
turbulent motions. Ultimately, these motions are what is powering the
redistribution of angular momentum on large scales. In the absence of
externally imposed large-scale fields, a separate dynamo mechanism may
be required to replenish sufficiently coherent fields to drive
sustained MRI turbulence \citep{2009ApJ...696.1021V}. This requirement
becomes particularly apparent in models that neglect the vertical
structure of the disk. In this case a convergence problem has been
encountered
\citep{2007ApJ...668L..51P,2007A&A...476.1113F,2011ApJ...739...82B},
which has been attributed to the lack of an outer scale of the
turbulence \citep{2010ApJ...713...52D}. An alternative explanation has
been suggested by \cite{2010A&A...513L...1K}, who point out the
problem of resolving the radial fine-structure related to
non-axisymmetric MRI modes (required to circumvent Cowling's ``no
dynamo'' theorem). Convergence can naturally be recovered when
accounting for the vertical stratification of the disk
\citep{2010ApJ...708.1716S,2011ApJ...740...18O}. Together with
rotational anisotropy, the introduced vertical inhomogeneity can
produce a pseudo-scalar leading to a classical mean-field dynamo
\citep{1995ApJ...446..741B}. In a previous paper \citep[hereafter
  \Gre]{2010MNRAS.405...41G}, we have inferred mean-field closure
parameters from isothermal stratified MRI simulations \citep[also
  see][]{2008AN....329..725B} and demonstrated that the ``butterfly''
pattern \citep[see
  e.g.][]{2000ApJ...534..398M,2010ApJ...708.1716S,2012MNRAS.422.2685S}
typical for stratified MRI can in fact be described in such a
framework. In the current paper, we aim to extend this line of work
towards a more realistic thermodynamic treatment, including heating
from turbulent dissipation
\citep{2005AIPC..784..475G,2009A&A...499..633P} and crude heat
transport. This new effort is largely initiated by a recent paper of
\citet*{2012ApJ...761..116B}, hereafter \Bodo. The authors make the
intriguing suggestion that the treatment of the disk thermodynamics
will have a strong effect on the dynamo and accordingly on the
saturation level of the turbulence. This is in contrast to the result
of \citet{1995LNP...462..385B}, who find that turbulent transport is
not affected by the presence of convection.

As a first step to improve the realism of MRI simulations compared to
the commonly applied isothermal equation of state, \Bodo assume
thermal conduction with constant thermal diffusivity, $\kappa$. They
then identify a critical value $\kappa_{\rm c}$ below which the
primary mode of vertical heat transport changes from being
predominantly conductive to being dominated by turbulent convective
motions. One motivation of the present work is to scrutinize these
models and at the same time obtain mean-field closure coefficients for
the two suggested regimes, thereby confirming the assumption that the
thermodynamic treatment of the disk affects underlying dynamo
processes. A second focus of our paper will be the authors' choice of
impenetrable vertical boundary conditions, which we find to have
profound implications for the resulting vertical disk equilibrium.


\section{Model and Equations}
\label{sec:methods}

The simulations presented in this paper extend the simulation of
\citet{2010MNRAS.405...41G} to include a more realistic treatment of
thermodynamic processes as pioneered by
\citet{2012ApJ...761..116B}. Simulations are carried out using the
second-order accurate \NIII code \citep{2004JCoPh.196..393Z}, which
has been supplemented with the HLLD Riemann solver
\citep{2005JCoPh.208..315M} for improved accuracy. We solve the
standard MHD equations in the shearing-box approximation
\citep{2007CoPhC.176..652G} employing the finite-volume implementation
of the orbital advection scheme as described in
\citet{2010ApJS..189..142S}; for interpolation we use the Fourier
method by \citet{2009ApJ...697.1269J}. We here neglect explicit
viscous or resistive dissipation terms but include an artificial mass
diffusion term \citep[as described in][]{2011MNRAS.415.3291G} to
circumvent time-step constraints due to low density regions in upper
disk layers. We remark that, owing to the total-variation-diminishing
(TVD) nature of our numerical scheme and the total energy formulation,
heating via dissipation of kinetic and magnetic energy at small scales
is accounted for even in the absence of explicit (or artificial)
dissipation terms.

Written in a Cartesian coordinate system ($\xx,\,\yy,\,\zz$) and with
respect to conserved variables $\rho$, $\rho\V$, and the \emph{total}
energy $e= \epsilon+\frac{1}{2}\rho\V^2 +\frac{1}{2}\B^2$, with
$\epsilon$ being the thermal energy density, the equations read:
\begin{eqnarray}
  \partial_t\rho +\nabla\cdt(\rho \V) & = & 0           \,,\\[6pt]
  \partial_t(\rho\V) +\nabla\cdt
          [\rho\mathbf{vv}+p^{\star}-\mathbf{BB}] & = &
          \rho\,[ -\nabla\Phi + \mathbf{a}_{\rm i} ]    \,,\\[6pt]
\label{eq:energy}
  \partial_t e + \nabla\cdt
          [(e + p^{\star})\V - (\V\cdt\B)\B] & = &
          \rho \V\cdt [ -\nabla\Phi + \mathbf{a}_{\rm i} ] \nonumber\\
    & + & \nabla\cdt(k\,\nabla T)                       \,,\\[6pt]
  \partial_t \B -\nabla\tms(\V\tms\B) & = & 0  \,,\nonumber\\
  \nabla\cdt\B & = & 0\,,
\label{eq:mhd}
\end{eqnarray}
with the total pressure given by $p^{\star}\equiv p+\frac{1}{2}\B^2$,
and a fixed external potential $\Phi(z)=\frac{1}{2}\Omega^2z^2$. The
inertial acceleration $\mathbf{a}_{\rm i} \equiv 2\Omega\,( q \Omega
x\,\xx - \zz\tms\V )$ arises due to tidal and Coriolis forces in the
local Hill system, rotating with a fixed $\OO\equiv\Omega_0\zz$, and
where the shear-rate $q\equiv {\rm d}\ln \Omega/{\rm d}\ln R$ has a
value of $-3/2$ for Keplerian rotation. 

We furthermore assume an adiabatic equation of state, such that the
gas pressure relates to the thermal energy density as
$p=(\gamma-1)\,\epsilon$, with the ratio of specific heats
$\gamma=\frac{5}{3}$, as appropriate for a mono-atomic dilute gas.

Finally, the temperature, $T$, appearing in the conductive energy
flux, is obtained via the ideal-gas law $p=\rho\,T$, where we chose
units such that the factor $\bar{\mu}\m_{\rm H}/k_{\rm B}$ relating to
the gas constant disappears. Following the approach taken by \Bodo, we
adopt a thermal conductivity, $k$, in terms of a constant diffusivity
coefficient $\kappa$, related via
\begin{equation}
  k = \frac{\gamma}{\gamma-1}\,\rho\,\kappa\,.
\end{equation}
For the isothermal run, we do not evolve \Eq{eq:energy} but instead
obtain the gas pressure via $p=\rho T_0$, with fixed temperature
$T_0=1$. Note that, in our units, $T_0$ differs from \Bodo by a factor
of two, owing the alternative definition of the initial hydrostatic
equilibrium, which is
\begin{equation}
  \rho(z) = \rho_0\, {\rm e}^{-{z^2}/{2H^2}} =
  \rho_0\, {\rm e}^{-{\Omega^2z^2}/{T_0}}
  \label{eq:hydrostat}
\end{equation}
in our case. For all simulations, we adopt the same box size of
$H\times\pi H\times 6H$ at a numerical resolution of $32\times
96\times 192$ grid cells in the radial, azimuthal, and vertical
direction, respectively. This corresponds to a linear resolution of
$\sim32/H$ in all three space dimensions.

As typical for shearing box simulations, we initialize the velocity
field with the equilibrium solution $\V=q\Omega\,x\yy$, and
adiabatically perturb the density and pressure by a white-noise of
$1\%$ rms amplitude. The magnetic configuration is of the
zero-net-flux (ZNF) type with a basic radial variation
\begin{equation}
\B=B_0\,\sin(2\pi x/L_x)\zz\,.
\end{equation}
To obtain a uniform transition into turbulence, we further scale the
vertical field with a factor $(p(z)/p(0))^{1/2}$ resulting in
$\mn{\beta}_{\rm P}={\rm const}$ ($=1600$ initially).\footnote{Here
  and in the following, over-bars denote horizontal
  averaging. Additional averaging in time is denoted by angle
  brackets, $\rms{\,}$.} Owing to the divergence constrain, this of
course can only be done by introducing a corresponding radial field at
the same time. In practice, we specify a suitable vector potential.

Horizontal boundary conditions (BCs) are of the standard
sheared-periodic type, and we correct the hydrodynamic fluxes to
retain the conservation properties of the finite-volume scheme
\citep{2007CoPhC.176..652G}. For the vertical boundaries, we implement
stress-free BCs (i.e., $\partial_z v_x=\partial_z v_y=0$) with two
different cases for the treatment of the vertical velocity component
$v_z$, namely: (i) impenetrable, and (ii) allowing for outflow (but
preventing in-fall of material from outside the domain). We will
demonstrate that this distinction will have profound implications for
the resulting density and temperature profiles within the box. To
counter-act the severe mass loss occurring in the case of open BCs, we
continuously rescale the mass density, keeping the velocity and
thermal energy density intact.\footnote{Technically, this amounts to a
  volumetric cooling of the disk.} Unlike in earlier work
\citep{2012MNRAS.422.1140G}, which was adopting an isothermal equation
of state, we here do not restore towards the initial profile, but
simply rescale the current profile. This is, of course, essential to
allow for an evolution of the vertical disc structure, owing to
heating via turbulent dissipation. We remark that such a replenishing
of material can be thought of as a natural consequence of radial mass
transport within a global disk.

As in \Bodo, we use $\partial_z B=0,\,B_x=B_y=0$ as boundary condition
for the magnetic field, and impose a constant temperature $T=T_0$ at
the top and bottom surfaces of the disk. The latter choice is
motivated by the assumption that the upper disk layers are likely
optically thin, and there exists a thermal equilibrium with their
surroundings. As in previous work, we compute the density and thermal
energy of the adjacent grid cells in the $z$~direction to be in
hydrostatic equilibrium.


\section{Results}
\label{sec:results}

The main motivation of this paper is to reproduce, as closely as
possible, the results of \Bodo, where we then aim to establish
mean-field dynamo effects for the contrasting cases of efficient
versus inefficient thermal conduction, as studied there. Moreover, we
shall begin to explore the impact of vertical boundary conditions, by
studying the somewhat more realistic case allowing for a vertical
outflow of material.

\begin{table}\begin{center}
\caption{Overview of simulation parameters, and results. \label{tab:models}}
\begin{tabular}{lccccc}
  \hline\hline
  & EoS & $\kappa\;[H^2\Omega]$ & vBC & $\rms{\mn{M}_{xy}}\;[10^{-2}p_0] $
  & $T_{\rm mid}\;[T_0]$
  \\[2pt]\hline
  M1 & isoth. & --      & outfl. & $0.54\pm 0.14$ & --   \\
  M2 & adiab. & $0.120$ & outfl. & $0.63\pm 0.17$ & 1.23 \\
  M3 & adiab. & $0.004$ & outfl. & $0.81\pm 0.21$ & 1.74 \\
  M4 & adiab. & $0.004$ & wall   & $1.53\pm 0.34$ & 4.94 \\
  \hline
\end{tabular}
\end{center}
\end{table}

We have performed in total four simulations: the two main simulations
(`M2' and `M3') adopt thermal diffusivity of $\kappa=0.12$, and
$\kappa=0.004$, respectively, representing the regimes of efficient
and inefficient thermal transport (at molecular level),
respectively. For both these simulations, and for a third isothermal
reference run (model `M1'), we adopt \emph{outflow} boundary
conditions. To assess the impact of the imposed vertical BCs (see
column ``vBC'' in \Tab{tab:models}), and to make direct contact with
previous work, we adopt a fourth model, `M4', with a value
$\kappa=0.004$, and \emph{impenetrable} boundaries at the top and
bottom of the domain (labeled ``wall'' in the following).


\subsection{Comparison with \Bodo}

\begin{figure}
  \center\includegraphics[width=0.95\columnwidth]{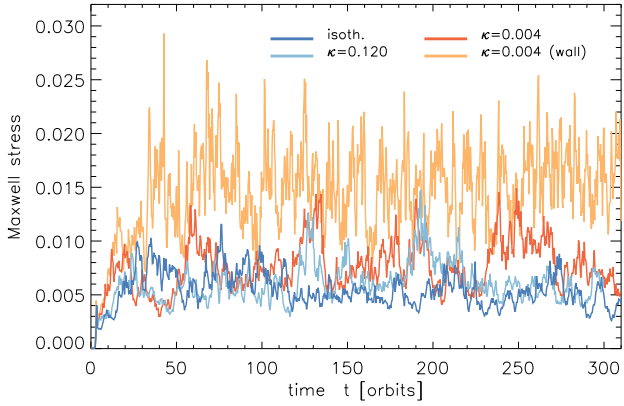}\\[1ex]
  \caption{Time evolution of the volume-averaged Maxwell stress for
    the different models. For reference, we also list time averages in
    \Tab{tab:models}.}
  \label{fig:time_Maxw}
\end{figure}

We ran the different models for approximately 300 orbital times,
$2\pi\Omega^{-1}$; note that \Bodo use time units of $\Omega^{-1}$
instead. All time averages are taken in the interval $t=[50,300]$. For
the isothermal reference run, we obtain a time-averaged Maxwell
stress,
\begin{equation}
\rms{\mn{M}_{xy}}\equiv -\rms{(B_x\!-\!\mn{B}_x)\;
  (B_y\!-\!\mn{B}_y)}\,,
\end{equation}
of $( 0.54\pm 0.14)\times 10^{-2}$, which is comparable to the
$\kappa=0.12$ case with $( 0.63\pm 0.17)\times 10^{-2}$. In the case
$\kappa=0.04$ with outflow boundaries we find a somewhat higher value
of $( 0.81\pm 0.21)\times 10^{-2}$, which is however significantly
lower than in the otherwise identical case with impenetrable
boundaries with $( 1.53\pm 0.34)\times 10^{-2}$. While stresses are
indeed increased (by approximately 30\%) at lower conductivity,
clearly, the effect of the treatment of the boundaries is much more
significant.

\begin{figure}
  \center\includegraphics[width=0.95\columnwidth]{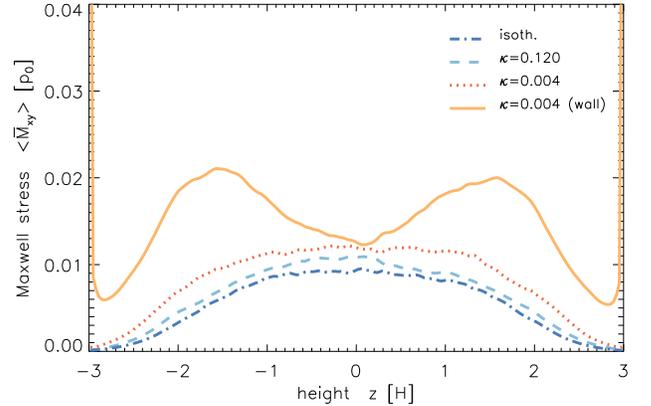}\\[1ex]
  \caption{Time-averaged (for $t\in[50,300]$ orbits) vertical profiles
    of the average Maxwell stress for the different models.}
  \label{fig:prof_Maxw}
\end{figure}

Due to the strong fluctuations, the relative amplitudes are best seen
in \Fig{fig:prof_Maxw}, where we plot time-averaged vertical profiles
of $\mn{M}_{xy}$ normalized to the initial midplane gas pressure,
$p_0$. Compared to the isothermal case M1, the thermally conductive
model M2 has a very similar vertical structure. The low conductivity
model M3, where turbulent overturning motions dominate, is not too
different in terms of its vertical profile either. Markedly, model M4,
with impenetrable vertical BCs, shows maxima of $\mn{M}_{xy}$ around
$|z|=1.5\,H$. This is similar to the profiles shown in figure~8 of
\Bodo, but note that their model shows a strong peak within $0.5\,H$
of the vertical boundaries, while we only observe a very thin boundary
layer in our model.

\begin{figure}
  \begin{center}
    \includegraphics[width=0.97\columnwidth]{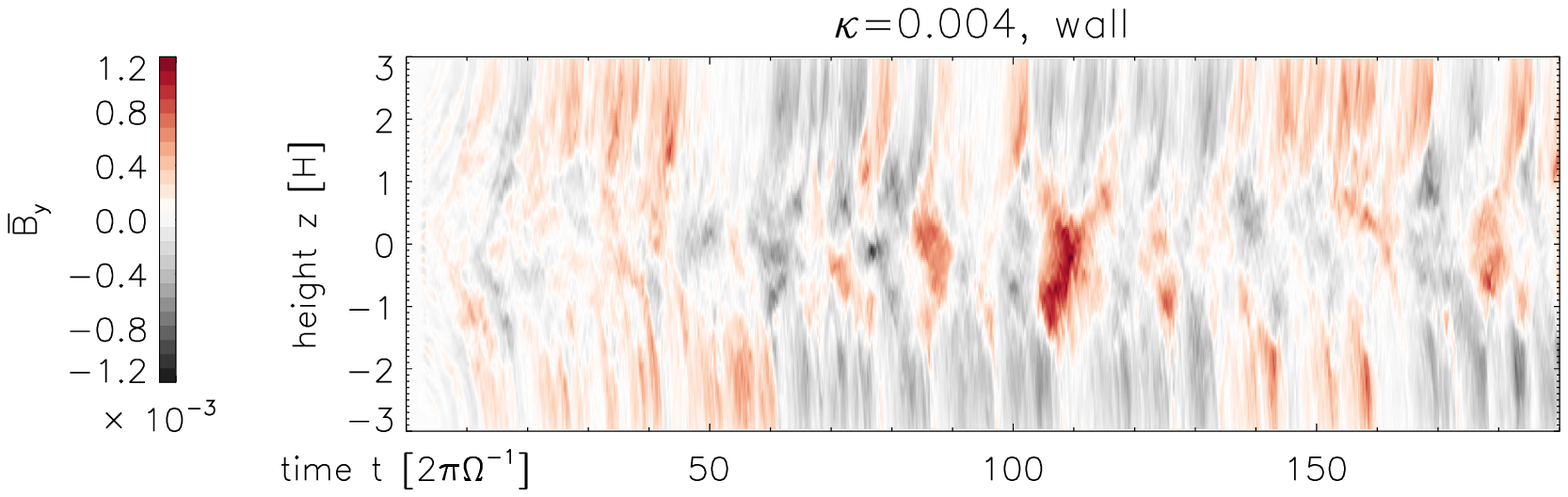}\\[2ex]
    \includegraphics[width=0.97\columnwidth]{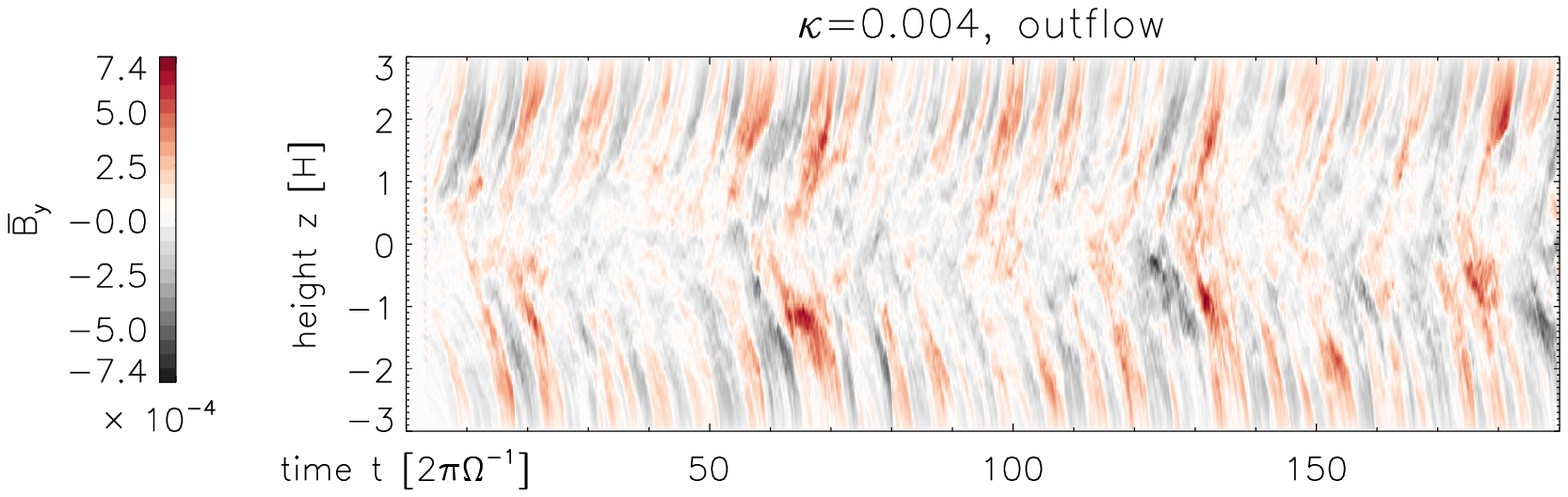}\\[2ex]
    \includegraphics[width=0.97\columnwidth]{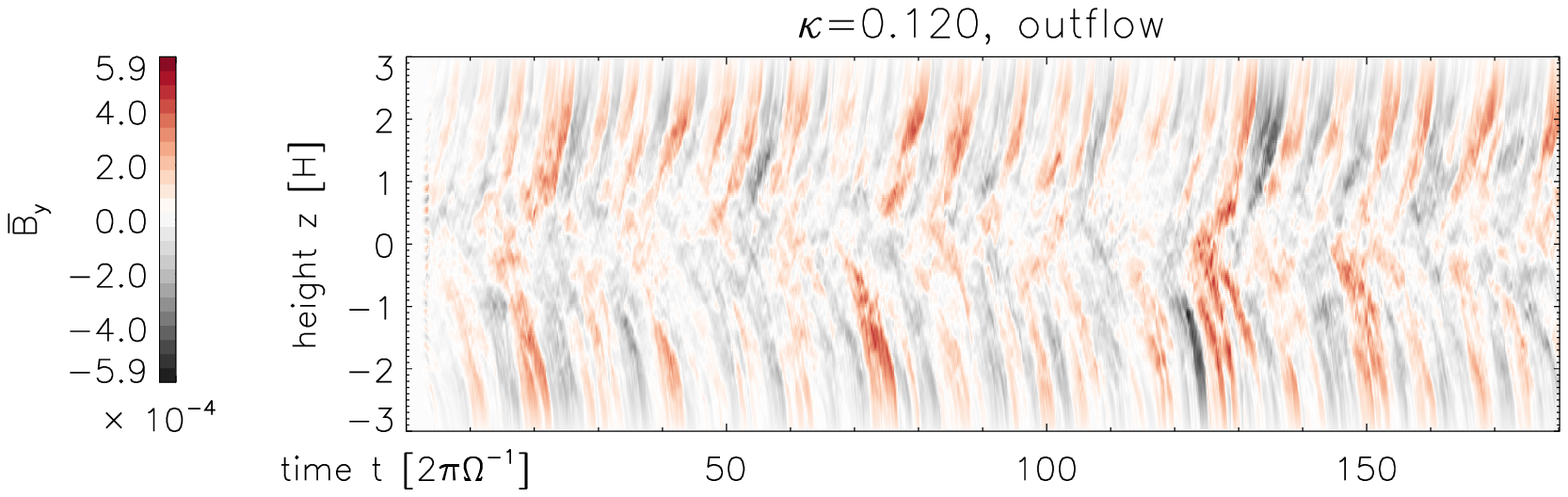}
  \end{center}
  \caption{Space-time ``butterfly'' diagrams of the azimuthal magnetic
    field $\mn{B}_y$ for: model M4 with $\kappa=0.004$ and
    impenetrable boundaries (top), model M3 with $\kappa=0.004$ and
    outflow boundaries (middle), and model M2 with $\kappa=0.12$ and
    outflow (lower panel).\\[1ex]}
  \label{fig:spctm}
\end{figure}

At the end of their result section, \Bodo point-out that the
spatio-temporal behavior of the dynamo is remarkably different between
the $\kappa=0.12$, and $\kappa=0.004$ cases (see their figure~9). If
we compare to run M4 with impenetrable boundaries (uppermost panel of
\Fig{fig:spctm}), we indeed find a similarly irregular butterfly
diagram. \Bodo argue that they find ``no evidence for cyclic activity
or pattern propagation'' in the conductive regime. In contrast to
this, looking at \Fig{fig:spctm}, it appears that even model M4 at
some level shows a propagating dynamo wave. More importantly, models
M2, and M3 (which presumably are in the conductive, and convective
regimes, respectively) show extremely similar dynamo patterns and
cycle frequency (middle and lower panels in \Fig{fig:spctm}). This
again suggests that the vertical boundaries have a profound effect on
the mechanism of vertical heat transport and, as a consequence, on the
dynamo.

\begin{figure}
  \center\includegraphics[width=\columnwidth]{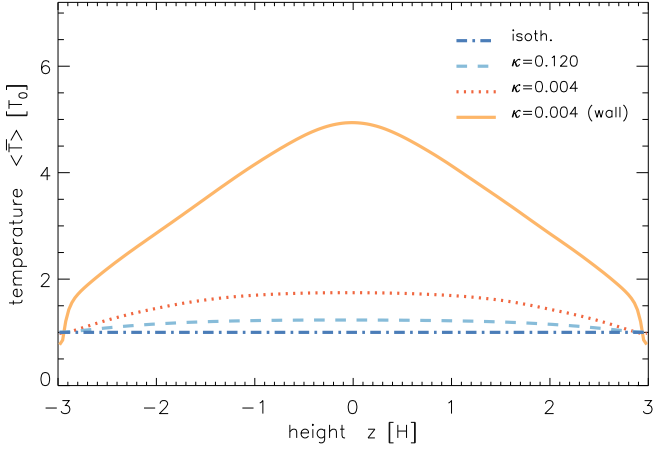}\\[1ex]
  \caption{Time-averaged (for $t\in[50,300]$ orbits) temperature profiles.}
  \label{fig:prof_T}
\end{figure}

To establish the fact that our model M4 is indeed comparable to the
corresponding simulation of \Bodo, we now look at vertical profiles of
the gas density and temperature. For the latter, \Bodo had found a
peculiar ``tent'' shape (see their figure~4), i.e. a linear dependence
of $\mn{T}$ on $z$, joined together at $z=0$ by a parabolic
segment. In \Fig{fig:prof_T}, we compile the $\mn{T}(z)$ profiles, and
such a tent-like profile can indeed be seen for model M4. However, in
the case of low thermal diffusivity, $\kappa=0.004$, we observe a much
weaker deviation from the isothermal profile with outflow boundary
conditions.

\begin{figure}
  \center\includegraphics[width=\columnwidth]{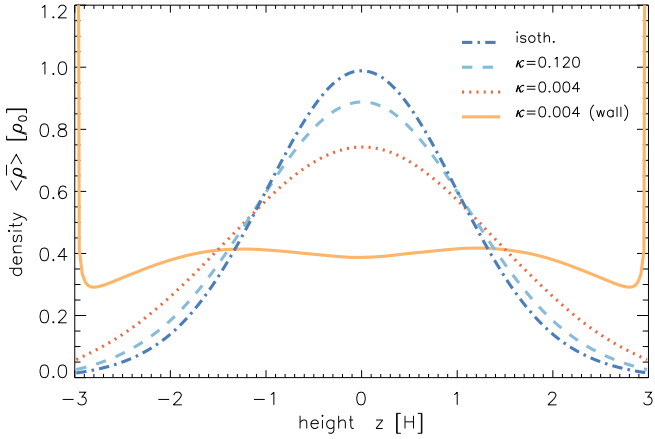}\\[1ex]
  \caption{Same as \Fig{fig:prof_T}, but for the average density.}
  \label{fig:prof_rho}
\end{figure}

Along with the tent-shaped temperature profile, \Bodo found that heat
transport from convection would erase the vertical density
stratification and lead to a constant density near the disk midplane
(see their figure~5). We observe a very similar density profile for
model M4, which moreover shows strong density peaks at the domain
boundary. These peaks are a consequence of enforcing the hydrodynamic
fluxes to be zero at the domain boundaries. Unlike for impenetrable
vertical boundaries, model M3 shows a more regular Gaussian density
profile. For models M1-M3, the width of the bell-shaped density
profiles is consistent with the trend in temperature and reflects
hydrostatic equilibrium. Apparently, such an equilibrium cannot be
obtained if solid-wall boundaries are applied.

\begin{figure}
  \center\includegraphics[width=\columnwidth]{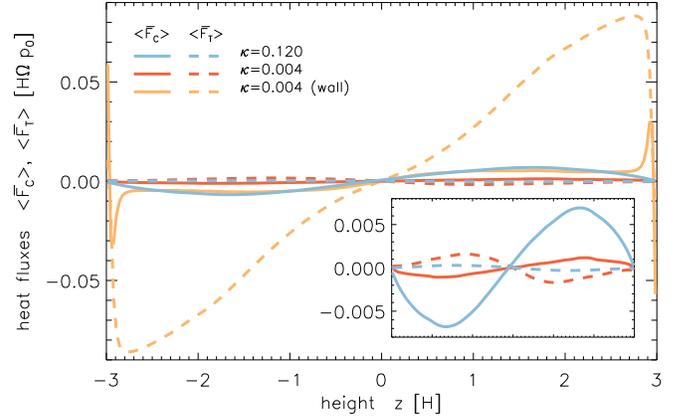}\\[1ex]
  \caption{Conductive heat flux $\rms{\mn{F}_{\rm C}}$ (solid lines),
    and turbulent convective heat flux $\rms{\mn{F}_{\rm T}}$ (dashed
    lines) for the three non-isothermal models.}
  \label{fig:heat_flux}
\end{figure}

\Bodo conjecture that when $\kappa$ crosses a critical value of
$\kappa_{\rm c}\simeq 0.02$, the vertical heat transport changes from
being mainly conductive to being predominantly convective. This was
illustrated by their figure~6, where they compared the conductive heat
flux
\begin{equation}
  \mn{F}_{\rm C} = -\frac{\gamma}{\gamma-1}\,
                    \kappa\mn{\rho}\,\frac{{\rm d}\mn{T}}{{\rm d}z}\,,
  \label{eq:flux_c}
\end{equation}
with the mean turbulent heat flux
\begin{equation}
  \mn{F}_{\rm T} = \frac{\gamma}{\gamma-1}\;
                   \overline{\rho v_z (T-\mn{T})} \,,
  \label{eq:flux_t}
\end{equation}
where horizontal lines as usual indicate averages over the $x$ and
$y$~directions. In \Fig{fig:heat_flux}, we show the corresponding
quantities for our non-isothermal runs: as in previous plots, model M4
agrees very well with the result of \Bodo, but again differs
significantly from the otherwise identical model M3 with outflow
boundary conditions. Compared to M4, the net transport of heat is much
reduced in the case of models M2, and M3. For clarity we plot these
curves separately in the inset of \Fig{fig:heat_flux} with magnified
ordinate. Much as expected, model M2 is dominated by a positive
(i.e. outward) conductive heat flux, whereas the turbulent flux is
negligible. Unlike model M4 with solid-wall boundaries, the
low-conductivity model, M3, only shows a very moderate level of
convective heat flux, and notably one of the opposite sign. This
negative flux appears to largely balance its positive conductive
counterpart, implying low levels of net-conductive heat transport
towards the disk surface and net-convective transport towards the
midplane. It is instructive to note that the heating predominantly
occurs at $z\simeq\pm 2\,H$, where the velocity and magnetic field
fluctuations are highest, and not, as one might think, near the
midplane.


\subsection{Mean-field coefficients}

One objective of this work was to test the dependence of the dynamo on
the mechanism of vertical heat transport. One way of doing this is to
establish mean-field closure parameters for the new class of models
with constant thermal diffusivity. Like in previous work, we utilize
the test-field (TF) method
\citep{2005AN....326..245S,2007GApFD.101...81S} to measure
coefficients such as the $\alpha$~effect, turbulent pumping, and eddy
diffusivity. The used method \citep{2005AN....326..787B} is
``quasi-kinematic'' \citep{2010A&A...520A..28R} in the sense that it
has been found to remain valid into the non-kinematic regime in the
absence of magnetic background fluctuations
\citep{2008ApJ...687L..49B} -- whether this covers MRI is a topic of
discussion. We here only briefly recapitulate the general framework of
mean-field MHD and refer the reader to \Gre for a more detailed
description. A recent review about mean-field dynamos can be found in
\citet{2012SSRv..169..123B}.

For the shearing-box approximation, due to its periodic character in
the horizontal direction, there are no characteristic gradients
expected in the radial or azimuthal directions. The natural
mean-fields are accordingly those, which only vary in the vertical
direction. With respect to the velocity $\U=\V-q\Omega x \yy$, the
mean-field induction equation reads
\begin{eqnarray}
  \partial_t \mB(z) & = \nabla \times & \left[\ \mU(z)\tms\mB(z)
    + \EMF(z) \right. \nonumber\\ 
    & & + \left. (q\Omega x\yy)\tms\mB(z)\ \right] \,, 
  \label{eq:MF_ind}
\end{eqnarray}
where we have ignored a contribution due to microscopic magnetic
diffusivity. Note that the explicit $x$~dependence in the shear term,
$q\Omega x \yy$, drops out once the curl operation is
applied. Furthermore, $\mn{B}_z = \mn{B}_z(t=0) = 0$ because of flux
conservation in the periodic box. In this description, turbulence
effects due to \emph{correlated} velocity and magnetic field
fluctuations are embodied in the mean electromotive force
\begin{equation}
  \EMF(z) \equiv \overline{\U'\tms \B'}\,,
\end{equation}
which is typically parametrized as \citep{2005AN....326..787B}:
\begin{eqnarray}
  \EMF_i(z) & = \alpha_{ij}(z)\ \mn{B}_j(z) &
         - \tilde{\eta}_{ij}(z)\ \varepsilon_{jkl}\,\partial_k \mn{B}_l(z)\,,
         \nonumber\\[4pt]
         & &  {\rm where\;} i,j \in \left\{x,y\right\},\, k\!=\!z\,.
  \label{eq:closure}
\end{eqnarray}
Given explicit knowledge of the rank-two $\alpha$ and $\tilde{\eta}$
tensors, this closure allows to formulate the mean-field induction
equation (\ref{eq:MF_ind}) in terms of mean quantities alone, leading
to the classical $\alpha\Omega$~dynamo description, first applied to
MRI turbulence by \citet{1995ApJ...446..741B}.

\begin{figure}
  \center\includegraphics[height=1.6\columnwidth]{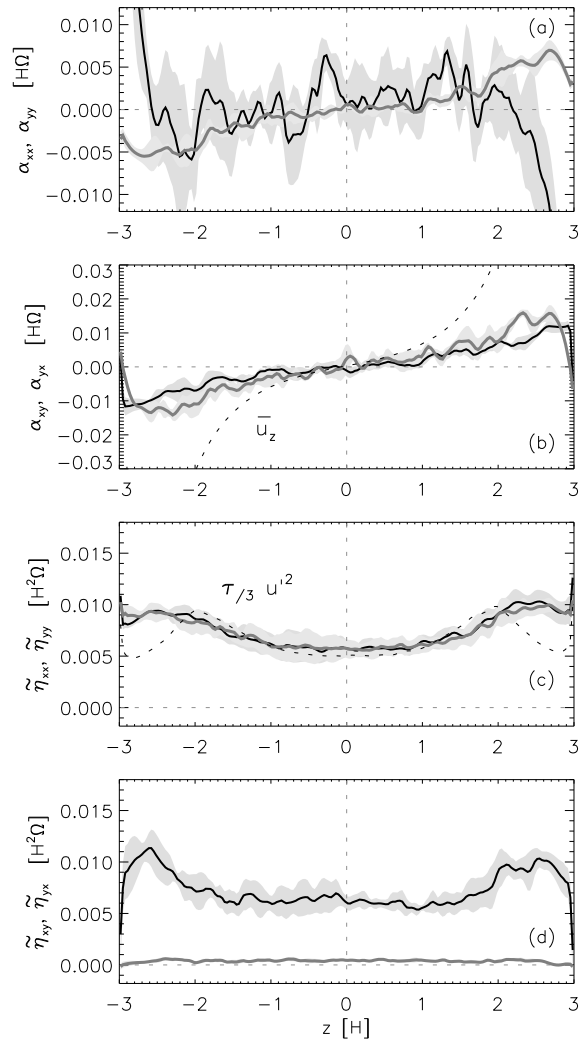}\\[1ex]
  \caption{Mean-field coefficients computed via the TF method for
    model M2 with $\kappa=0.120$. Axis labels indicate curves plotted
    in dark ($\alpha_{xx},\dots$) or light ($\alpha_{yy},\dots$)
    colours, respectively. The mean vertical velocity, $\mn{u}_z$, in
    panel~(b), and the classical estimate for the turbulent diffusion,
    in panel~(c), are shown as dashed lines.}
  \label{fig:dyn_k120}
\end{figure}

Tensor coefficients representing the closure parameters are presented
in \Fig{fig:dyn_k120} for model M2, with a high value $\kappa=0.12$ of
the thermal diffusivity. The obtained profiles are largely similar to
the corresponding curves of the isothermal model M1 (not shown), which
moreover agree\footnote{This is with the exception of $\alpha_{xx}$,
  which generally appears to be poorly constrained by our
  simulations.} with previous isothermal results reported in
\Gre. Notably, for model M2, the contrast between the disc midplane
and the upper disc layers is somewhat less pronounced compared to the
purely isothermal case. This trend is continued when going to lower
thermal diffusivity (see \Fig{fig:dyn_k004} below). Unlike reported in
\Gre, we now find $\alpha_{xx}$ and $\alpha_{yy}$ to be predominantly
of the \emph{same} sign, which would argue in favor of a kinematic
(rather than magnetic) origin of the effect. Note however the
significant fluctuations in $\alpha_{xx}$, which cast some doubt on
whether this coefficient can be meaningfully determined in the
presence of shear. On the other hand, a fluctuating $\alpha$ should
not be disregarded as a possible source of a mean-field dynamo
\citep{1997ApJ...475..263V}.

Before we proceed, we briefly discuss the remaining coefficients. In
panel~(b) of \Fig{fig:dyn_k120}, we show the off-diagonal tensor
elements of the $\alpha$~tensor, which are dominantly symmetric,
i.e. $\alpha_{xy}\simeq\alpha_{yx}$. We remark that for the classical
diamagnetic pumping effect, one would require \emph{anti}-symmetric
parity. The observed symmetry may however be interpreted as
differential pumping, i.e. transporting radial and azimuthal field in
opposite directions. For reference, we plot the mean vertical velocity
$\mn{u}_z$ (see dashed line), which additionally transports the mean
field and hence leads to the characteristic acceleration in the
butterfly diagram. The diagonal parts of the $\tilde{\eta}$ tensor are
shown in panel~(c), where we also plot the rms velocity fluctuation
(dashed line). Apart from the boundary layers, the turbulent
diffusivity agrees well with the theoretical expectation
\begin{equation}
  \eta_{\rm T} \simeq \frac{\tau}{3}\,u'^{\,2}\,,
\end{equation}
that is, assuming a coherence time $\tau=0.03\,\Omega^{-1}$ of the
turbulence. Given the dominance of the azimuthal field, the
coefficients $\tilde{\eta}_{xx}$, and $\tilde{\eta}_{yy}$ are
surprisingly isotropic \citep*[unlike predicted
  by][]{1993ApJ...404..773V}.  It is however interesting to note that
while $\tilde{\eta}_{xy}$, shown in panel~(d) of \Fig{fig:dyn_k120},
is identical to the diagonal elements of the diffusivity tensor in
panel~(c), its counterpart $\tilde{\eta}_{yx}$ is much smaller. With
negative shear and both coefficients positive, the dynamo based on the
\citet{1969MDAWB..11..272R}~effect is decaying \citep[cf. the
  dispersion relation in][appendix~B]{2005AN....326..787B}. This does
however not exclude the possibility that $\tilde{\eta}_{yx}$ has an
effect on the overall pattern propagation.

\begin{figure}
  \center\includegraphics[height=1.6\columnwidth]{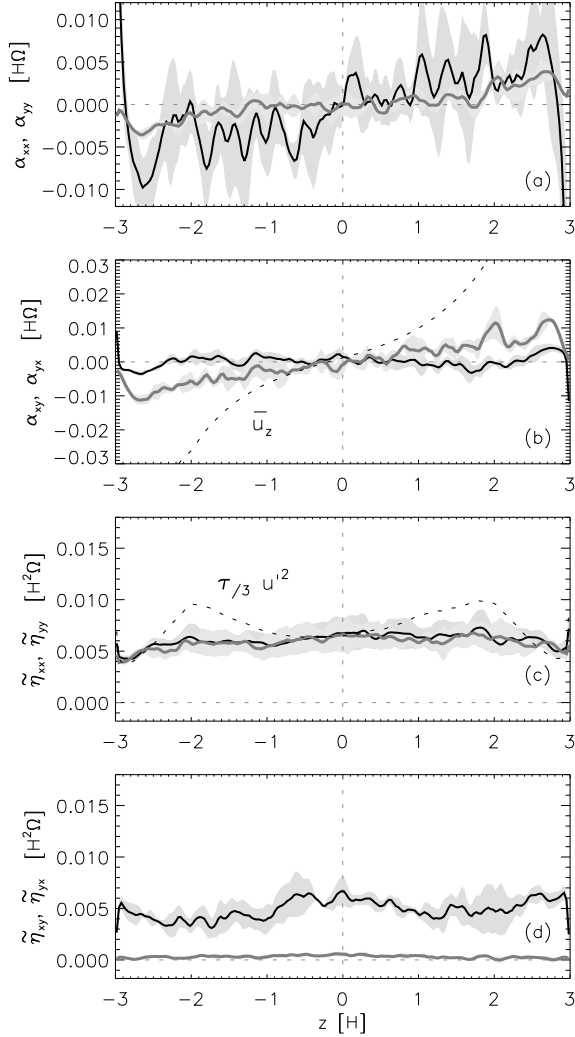}\\[1ex]
  \caption{Same as \Fig{fig:dyn_k120}, but for the model M3 with
    $\kappa=0.004$. Note that for the sake of direct comparison, axis
    ranges are kept fixed with respect to the previous figure.}
  \label{fig:dyn_k004}
\end{figure}

\begin{figure}
  \center\includegraphics[height=1.6\columnwidth]{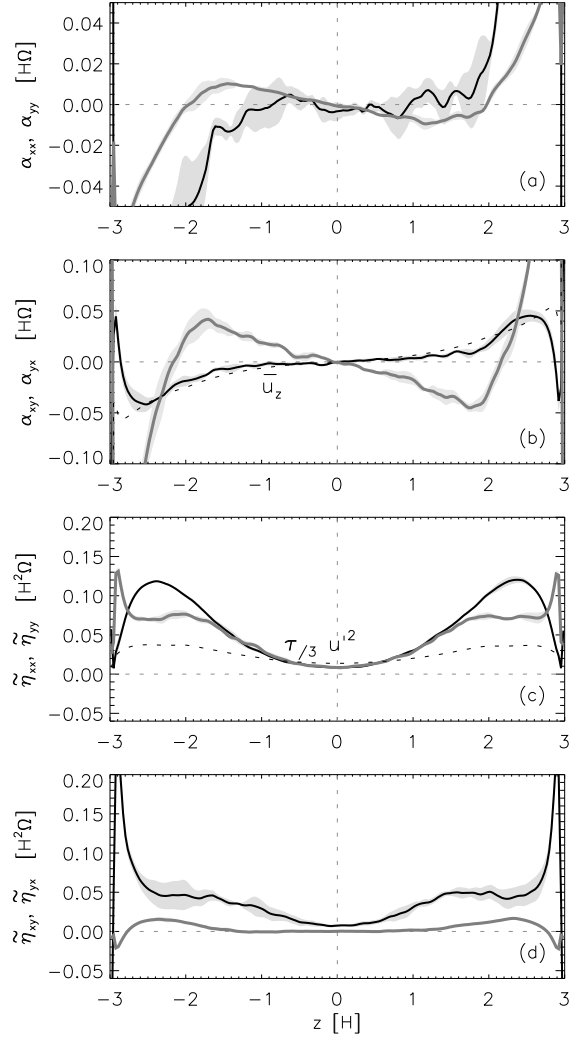}\\[1ex]
  \caption{Same as \Fig{fig:dyn_k120}, but for the model M4 with
    $\kappa=0.004$ and impenetrable vertical boundaries. Note the
    different axis ranges as compared to the previous two figures.}
  \label{fig:dyn_k004_wall}
\end{figure}

We now proceed to the corresponding coefficients for the case of
inefficient thermal conduction. In \Fig{fig:dyn_k004} we accordingly
show the $\alpha$ and $\tilde{\eta}$ tensor components for model
M3. We recall that in the quasi-isothermal case
(cf. \Fig{fig:dyn_k120}) the $\alpha_{xx}$ and $\alpha_{yy}$
coefficients showed a trend to flatten and even reverse their slope
near the midplane \citep[also cf. figure~9 in][]{2008AN....329..725B}.
This was reasoned to be related to a \emph{negative} $\alpha$~effect
due to magnetic buoyancy
\citep{1998tbha.conf...61B,2000A&A...362..756R}. In contrast, here the
$\alpha_{xx}$ curve shows a more monotonic dependence on $z$,
indicating that such magnetic effects may be less pronounced in this
case. Such a trend appears consistent with reduced magnetic buoyancy
in the case of a stiffer effective equation of state.  Moving on to
panel~(b), we note that $\alpha_{xy}$ is now suppressed and even shows
a slight trend to change its sign -- indicating a possible
significance of diamagnetic pumping in the adiabatic case. For the
turbulent diffusivity plotted in panel~(c), we observe a significant
deviation from the classical expectation (dashed line), resulting in a
nearly constant $\eta_{\rm T}(z)$. We conclude that the equation of
state and the means by which energy is transported to the upper disk
layers indeed have subtle effects on the inferred dynamo
tensors. Which of the differences seen between
\Figs{fig:dyn_k120}{fig:dyn_k004} is in the end responsible for the
enhanced dynamo activity seen in model M3, will require further
careful study. For completeness, in \Fig{fig:dyn_k004_wall}, we also
show the results for model M4, where dynamo coefficients are larger by
a factor of several. This can be considered consistent with the much
higher Maxwell stresses observed in this case. Unlike for model M3,
$\tilde{\eta}(z)$ now shows a pronounced $z$~dependence, and
$\alpha_{yy}$, and $\alpha_{yx}$ are in fact
negative\footnote{i.e. for $z>0$, and accordingly positive for $z<0$}
(for $|z|<2\,H$) as suggested by \citet{1998tbha.conf...61B}.

\section{Conclusions}
\label{sec:concl}

The primary goal of our work was to make contact with recent results
by \Bodo, and to complement their work with a direct measurement of
mean-field dynamo effects via the TF method. Given that we have used a
very similar numerical method and applied identical parameters
(i.e. for model M4), it should not surprise the reader that we can
satisfactorily confirm all aspects of the corresponding simulation by
\Bodo. Minor discrepancies arise with respect to the boundary layers,
which may be related to the detailed treatment of the hydrostatic
equilibrium there.

Accordingly, we can confirm their main result, namely that -- in the
presence of \emph{impenetrable} vertical boundary conditions -- one
observes a transition from a conductively dominated vertical heat
transport to a state that is regulated by convective overturning
motions. This transition obviously depends on the value of the applied
constant thermal diffusivity $\kappa$. Like reported in \Bodo, in the
$\kappa=0.004$ case, we observe a flat density profile (even with a
slight minimum at the disk midplane), and a ``tent''-shaped
temperature profile -- presumably established by convective heat
transport as a result of Rayleigh-Taylor-type instability. This case
is also associated with a much increased dynamo activity (by an order
of magnitude in the TF coefficients, not shown here), resulting in an
overall Maxwell stress that is increased by a factor of three compared
to the isothermal reference model. Unlike speculated by \Bodo, we
however do not think that the enhanced dynamo activity is related to
the magnetic boundary conditions. This is despite the fact that such a
connection indeed exists for the unstratified case
\citep{2011MNRAS.413..901K}, where the different magnetic boundary
conditions serve to create an inhomogeneity in an otherwise
translationally symmetric system. We rather attribute the different
dynamo regime to the overall different hydrodynamic state -- which
however appears largely influenced by the choice of impenetrable
boundary conditions.

A separate set of models (M1-M3) with a more natural condition
allowing the gas to flow out of the domain shows much less dramatic
effects when going to the low thermal diffusivity regime. Naturally,
one arrives at moderately hotter disk interiors along with more
spread-out, yet still Gaussian density profiles. Dynamo TF
coefficients are somewhat altered in this case, along with a roughly
30\% higher turbulent Maxwell stress. Establishing a link
\citep{2010AN....331..101B} between the turbulent transport
coefficients in the momentum equation (i.e. the Maxwell and Reynolds
stresses) and the induction equation (i.e. the $\alpha$~effect,
turbulent diffusion, etc.) will be key to understand magnetized
accretion in a quantitative manner, and derive powerful closure models
in the spirit of \citet{2003MNRAS.340..969O} or
\citet{2006PhRvL..97v1103P}. A possible direct extension of the
existing models with varying amounts of thermal conductivity may be to
cross-correlate $\rms{\mn{M}_{xy}}$ with e.g. $\alpha_{yy}$ for various
values of $\kappa$. Such a connection has been suggested by
\citet{1998tbha.conf...61B} and been derived in the quasi-linear
regime by \citet{2000A&A...362..756R}. A complication in this endeavor
however arises from the fact that the TF coefficients are likely
measured in a magnetically affected, i.e. quenched state.

Given the dramatic effect of open versus closed vertical boundaries
demonstrated in this paper, it will be of prime interest to study the
connection between the disk and the launching of a magnetically driven
wind \citep{2012MNRAS.423.1318O}, including a possible influence of
the wind on the disk dynamo. The amount of recent work
\citep{2012arXiv1210.6664F,2012A&A...548A..76M,%
  2012arXiv1210.6661B,2013A&A...550A..61L} illustrates the importance
of this issue. In terms of the vertical boundaries imposed on the
temperature, including a transition into an optically-thin disk corona
will be important. Then a radiative boundary condition consistent with
black-body radiation can be applied. To conclude, we want to emphasize
that, clearly, the presented simulations can only be regarded as a
first step towards a realistic treatment of the disk
thermodynamics. Ideally, full-blown radiative transfer should be
employed, and simulations of radiation dominated accretion disks
\citep{2011ApJ...733..110B} demonstrate that this has indeed become
feasible.

\acknowledgments 

The author acknowledges the anonymous referee for providing a
well-informed report and wishes to thank Axel Brandenburg and
Gianluigi Bodo for useful comments on an earlier draft of this
manuscript. This work used the \NIII code developed by Udo Ziegler at
the Leibniz Institute for Astrophysics (AIP). Computations were
performed on resources provided by the Swedish National Infrastructure
for Computing (SNIC) at the PDC Centre for High Performance Computing
(PDC-HPC).





\end{document}